\documentclass[conference]{IEEEtran}
\IEEEoverridecommandlockouts

\usepackage[a4paper,top=19mm,bottom=43mm,left=13mm,right=13mm,columnsep=6.3mm]{geometry}
\usepackage{cite}
\usepackage{url}
\usepackage{amsmath,amssymb,amsfonts}
\usepackage{graphicx}
\usepackage{booktabs}
\usepackage{enumitem}
\usepackage{algorithm}
\usepackage{algpseudocode}
\usepackage{ifthen}
\usepackage{seqsplit}
\usepackage[most]{tcolorbox}
\usepackage{textcomp}
\usepackage{xcolor}
\usepackage[hidelinks]{hyperref}
\usepackage{wrapfig}
\setlist[itemize]{noitemsep, topsep=2pt}

\newif\ifcomment
\commenttrue
\newboolean{showcomments}
\setboolean{showcomments}{true}

\newcommand{\yw}[1]{{\ifthenelse{\boolean{showcomments}} {\color{olive}{#1}}{}}}
\newcommand{\skr}[1]{{\ifthenelse{\boolean{showcomments}} {\color{blue}{#1}}{}}}
\newcommand{\yf}[1]{{\ifthenelse{\boolean{showcomments}} {\color{red}{#1}}{}}}

\commentfalse

\def\BibTeX{{\rm B\kern-.05em{\sc i\kern-.025em b}\kern-.08em
    T\kern-.1667em\lower.7ex\hbox{E}\kern-.125emX}}

\newif\ifanonymous
\anonymousfalse

\begin{document}

\bstctlcite{IEEEtran:BSTcontrol}

\title{BridgeHammer: DNN Backdoor Attack via CPU-iGPU Shared-memory Rowhammer}
\ifanonymous
\author{Anonymous Author(s)}
\else
\author{
\IEEEauthorblockN{Yufei Wang}
\IEEEauthorblockA{\textit{Electrical and Computer Engineering}\\
\textit{Northeastern University}\\
Boston, MA, USA\\
wang.yufei1@northeastern.edu}
\and
\IEEEauthorblockN{Saion K. Roy}
\IEEEauthorblockA{\textit{Electrical and Computer Engineering}\\
\textit{Northeastern University}\\
Boston, MA, USA\\
sai.roy@northeastern.edu}
\and
\IEEEauthorblockN{Yunsi Fei}
\IEEEauthorblockA{\textit{Electrical and Computer Engineering}\\
\textit{Northeastern University}\\
Boston, MA, USA\\
y.fei@northeastern.edu}
}
\fi

\maketitle

\begin{abstract}
Heterogeneous systems-on-chips (SoCs) share physical memory between CPUs and integrated GPUs (iGPUs), creating a path for CPU-side memory corruption to compromise the integrity of deep learning models used for iGPU inference. Implanting a backdoor that causes targeted misclassifications on trigger-bearing inputs while retaining accuracy on normal inputs requires carefully selected weight changes. However, realizing these changes presents two challenges: where to induce faults, since vulnerable DRAM locations may not contain the targeted weights, and when to intervene, since memory placement and fault injection must align with different stages of the model deployment lifecycle that involve two distinct computation domains. We present BridgeHammer, a Rowhammer-based backdoor attack that addresses these challenges by coordinating targeted weight placement before zero-copy binding for the iGPU and fault injection before inference. The iGPU then uses the corrupted model for subsequent inferences. We evaluate BridgeHammer on a heterogeneous Intel CPU–iGPU platform using OpenVINO, targeting ResNet-20, VGG-16, and ResNet-50 at both INT8 and FP32 precision. Across all six configurations, BridgeHammer achieves an 88.9\% attack success rate (ASR) and 86.3\% clean accuracy on average, with reliable target-page placement and bit-flip reproduction. These results demonstrate a new integrity vulnerability on heterogeneous CPU-iGPU SoCs, 
motivating cross-domain memory integrity protections.

\end{abstract}


\section{Introduction}\label{sec:intro}

Deep neural networks (DNNs) are increasingly deployed on heterogeneous Systems-on-Chips (SoCs) that couple CPUs with integrated GPUs (iGPUs) under a unified memory architecture~\cite{surveypaper, EdgeNN, CoDL}. This shared-memory organization differs from the memory system of computing platforms with discrete GPUs, where model parameters are copied from host DRAM into the GPU’s private video memory (VRAM) for execution. Physically isolated from any adversary on the host CPU, the GPU-resident model preserves its integrity during inference. On integrated platforms, sharing the physical memory enables zero-copy DNN deployments in which the iGPU accesses weights of the model loaded into memory by the host, avoiding explicit host-to-device copies. This shared-memory execution path raises an integrity concern: \textit{an adversary on the host CPU can corrupt the memory to affect the model consumed by the iGPU for inference.}

Rowhammer poses a run-time threat to DNN integrity by inducing bit-flips in DRAM rows through repeated accesses to adjacent rows~\cite{Blacksmith, TRRespass}. An attacker can use these faults to implant a backdoor, causing targeted misclassifications on triggered inputs while preserving clean accuracy on normal inputs~\cite{dontknock, li2025oneflip}. Such an attack requires changes to specific model weights, but the memory locations susceptible to Rowhammer may not contain those weights. Realizing the intended corruption therefore requires placing the targeted weight pages at physical memory locations with compatible fault patterns.

Meeting this placement requirement is difficult through the interfaces used by existing GPU-side Rowhammer techniques. Although prior work has demonstrated bit-flips from integrated GPUs~\cite{grandpwaningunit, GPU-rowhammer-new-work}, these attacks operate through opaque graphics interfaces that do not directly expose physical-address mappings. Applying them to a targeted DNN backdoor therefore requires additional mechanisms to associate target weights with physical memory locations and even control the placement, which are nearly impossible. Because the CPU and iGPU share the physical memory, another approach is to perform Rowhammer from the CPU-side to corrupt model weights used by the iGPU.

Integrity attacks on DNN models via Rowhammering have been demonstrated on CPU platforms only, including DeepHammer~\cite{Deephammer}, OneFlip~\cite{li2025oneflip}, Don't Knock~\cite{dontknock}, and ROBBIN~\cite{robbin},  where the adversary and the victim model co-reside on the CPU. They all include a common memory massaging step, which manipulates the physical page frame allocation through the OS so that targeted model pages are placed in Rowhammer-vulnerable DRAM locations. They differ in the attack goal (accuracy reduction or backdoor) and the attack efficiency (how many bit-flips to incur on which weights). These techniques provide a basis for selecting and realizing the required placement. Applying them to iGPU inference introduces a major timing challenge: \textit{when to intervene in the model deployment lifecycle}.

On CPU-iGPU platforms, the model loading process typically runs on the CPU: the host CPU loads model parameters into physical DRAM page frames allocated by the OS, accessible through virtual addresses. The inference framework includes GPU-specific stages: it compiles the model into an executable form for the iGPU and sets up zero-copy access to the loaded weights before starting inference. The attacker’s actions must align with different stages of this sequence: memory massaging must occur after model compilation and before zero-copy weight anchoring, while Rowhammer must occur after weight anchoring and before inference dispatch. The attacker must therefore closely track the victim’s progress and coordinate two separate interventions, ensuring the targeted weights occupy the intended physical locations and are corrupted at the right time. 

In this work, we present BridgeHammer, a cross CPU--iGPU Rowhammer-based backdoor attack that coordinates weight placement and fault injection with the DNN deployment lifecycle. Given DRAM fault profiles and an offline-constructed backdoor~\cite {robbin}, BridgeHammer identifies two breakpoints where the attacker temporarily pauses victim execution and injects two deliberate functions: targeted memory massaging and Rowhammering. We implement BridgeHammer on an Intel Core i7-8700 running OpenVINO and evaluate it on ResNet-20, VGG-16, and ResNet-50 under INT8 and FP32 precision.

This work makes the following contributions:
\begin{itemize}[noitemsep,topsep=2pt,leftmargin=15pt]
\item We demonstrate a cross CPU--iGPU Rowhammer-based backdoor attack in which an adversary on the CPU corrupts model weights precisely to implant a backdoor in the model and affect subsequent iGPU inferences.
\item We characterize the CPU-to-iGPU deployment lifecycle and coordinate targeted weight placement and fault injection at two distinct stages, via instruction monitoring and signal-based synchronization. %
\item We evaluate BridgeHammer on an Intel platform running OpenVINO across three DNN models under INT8 and FP32. Across all six configurations, the attack achieves 100\% accurate target-page placement and bit-flip reproduction, with an 88.9\% attack success rate and 86.3\% clean accuracy on average.
\end{itemize}

The remainder of this paper is organized as follows. Section~\ref{sec:background} provides background. Section~\ref{sec:attack-meth} elaborates on our proposed attack methodology. Section~\ref{sec:attack-inst} describes the attack implementation on the Intel platform, followed by Section~\ref{sec:eval}, which presents the evaluation. Finally, Section~\ref{sec:disc} discusses security implications, and concludes the paper.

\section{Background and Related Work}\label{sec:background}

This section surveys prior backdoor attacks on DNN models and the Rowhammer primitives that enable them. We compare existing Rowhammer implementations on CPUs and GPUs, highlighting a research gap for integrated CPU-GPU platforms and shaping our choice of CPU-side Rowhammer strategy.

\subsection{Backdoor Attacks on DNN Models}

A backdoored DNN model produces attacker-chosen outputs when a trigger pattern is present in the input while behaving normally on benign inputs. The trigger is a small predefined perturbation applied to an input. The effectiveness of a backdoor is commonly measured by the attack success rate (ASR) on trigger-bearing inputs and the clean accuracy (CA) on benign inputs. A successful attack therefore aims to achieve high ASR while keeping CA close to the original model's baseline accuracy.

Prior backdoor attacks show that modifying a small number of carefully selected weight bits can implant such behavior. At the algorithm level, TBT~\cite{rakin2020tbt} demonstrates trigger-activated backdoor injection through targeted weight modifications, while ProFlip~\cite{chen2021proflip} reduces the required number of flipped bits by searching for the most backdoor-effective parameters. OneFlip~\cite{li2025oneflip} pushes this efficiency to an extreme, showing that a single FP32 bit-flip can be sufficient to install a backdoor on some models. Don't Knock~\cite{dontknock} incorporates hardware constraints, device-specific Rowhammer-induced faults, into backdoor construction for INT8 models by restricting target backdoor bit-flips to be sparse, one per page, and choosing vulnerable DRAM frames that can support the bit-flip while enduring side-effects of collateral unwanted bit-flips. %
More recently, ROBBIN~\cite{robbin} accounts for the complete device-specific fault patterns associated with candidate DRAM pages, and selects DNN-to-DRAM page mappings to achieve high backdoor ASR while preserving clean accuracy.

The threat model for all prior backdoor attacks differs from ours. They are CPU-only, where the adversary and victim DNN co-reside on the same CPU processor.  %
For our targeted SoCs with integrated CPU-iGPU where the DNN victim runs on the GPU, the model deployment pipeline crosses two domains: model weights are loaded and managed by the CPU but later consumed by the iGPU through zero-copy binding. The feasibility of realizing a backdoor attack through Rowhammer across this CPU-iGPU deployment lifecycle has not been investigated. BridgeHammer addresses this research gap.

\subsection{GPU-Side Rowhammer and Its Limits on Integrated Platforms}
 
Grand Pwning Unit~\cite{grandpwaningunit} demonstrates Rowhammer from integrated GPUs, while GPUHammer~\cite{gpuhammer} targets discrete GPU memory. Not So Refreshing~\cite{notsorefreshing} studies timing leakage involving in-DRAM Refresh Management (RFM). A targeted DNN backdoor also requires control over where the selected model weights reside. GPU interfaces obscure the physical-address mappings needed for that placement. BridgeHammer therefore uses CPU-side memory management and Rowhammer, with no access to the GPU execution interface.

\begin{figure*}[t]
    \centering
    \includegraphics[width=0.95\linewidth]{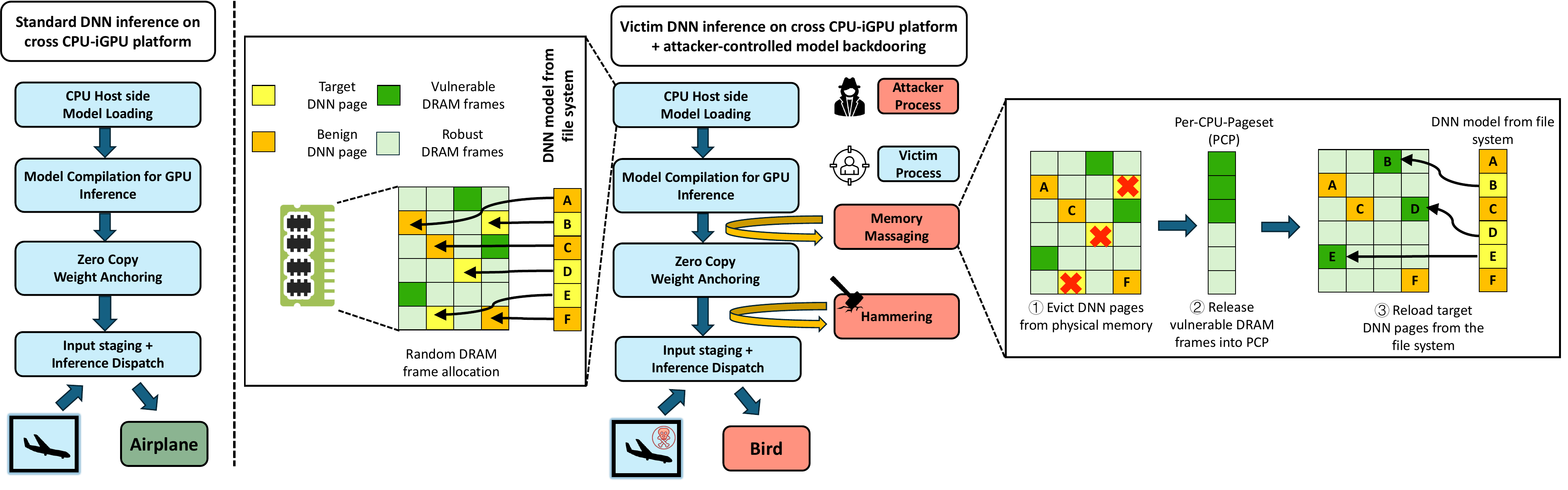}
    \caption{Left: Standard DNN deployment pipeline on CPU-iGPU platforms; Right: BridgeHammer online phase, the proposed cross CPU-GPU Rowhammer-based backdoor attack process}
    \label{fig:main-fig}
    \vspace{-0.5cm}
\end{figure*}

\section{Attack Methodology}\label{sec:attack-meth}

BridgeHammer consists of an offline phase and an online phase. The offline phase profiles the specific DRAM hardware to obtain its Rowhammer fault patterns and constructs the cognizant backdoor for a target DNN model. We adopt the same approach as prior work ROBBIN~\cite{robbin}. The outcome is a DNN pages-to-DRAM frames mapping for the backdoor on the victim system. The online phase then implants the backdoor in the model used in an ML service. The adversary interrupts the victim at two breakpoints to realize these mappings and fault patterns across the CPU-iGPU execution lifecycle. At the first breakpoint, the attacker performs memory massaging to place those target DNN pages on the suitable vulnerable DRAM frames. At the second breakpoint, the adversary performs Rowhammering on those vulnerable DRAM frames to introduce bit-flips needed by the backdoor. The hidden backdoor then persists across all subsequent inferences.  

The attack faces two challenges: precisely mapping \textit{target DNN pages} to vulnerable DRAM frames, and choosing appropriate breakpoints and synchronizing the attacker and victim. The remainder of this section describes the threat model and the attack methodology addressing the challenges.

\subsection{Threat Model}\label{subsec:attack-meth-model}

We consider a DNN inference server running on a heterogeneous CPU-iGPU platform with unified host-device memory. A pre-trained model is loaded from persistent storage, and the service waits for client queries to dispatch to the iGPU for processing with the deployed model. A co-located attacker runs a separate user-space process on the same machine. The attacker knows the publicly available target model and the DNN service's software stack. The attacker's capabilities include profiling the target DRAM for device-specific vulnerabilities and performing Rowhammering procedures, consistent with prior Rowhammer-based DNN attacks~\cite{Deephammer, dontknock, chen2021proflip}.

We further assume the attacker can observe the victim's execution progress through micro-architectural side channels and temporarily interrupt its execution. The attacker cannot modify the victim application, and does not have access to the iGPU execution interface. An effective backdoor attack involves four abstractions for the model and the DRAM. At the virtual memory level, the victim's model pages (data structures) are classified into:

\begin{itemize}[noitemsep,topsep=2pt,leftmargin=15pt]
    \item \textit{Target DNN pages}: the pages containing those parameters selected for corruption/bit-flips  prescribed by the backdoor.
    \item \textit{Benign DNN pages}: the remaining model pages whose integrity remains intact to maintain model clean accuracy.
\end{itemize}

\noindent At the physical memory level, we distinguish:

\begin{itemize}[noitemsep,topsep=2pt,leftmargin=15pt]
    \item \textit{Vulnerable DRAM frames}: the physical DRAM frames on rows that are susceptible to Rowhammer, and the induced bit-flips match what are required on \textit{target DNN pages} by the backdoor. 
    \item \textit{Robust DRAM frames}: physical pages that are neither targeted by Rowhammer nor on the aggressor rows which are physically adjacent to \textit{vulnerable DRAM frames} and used by the adversary.
\end{itemize}

\subsection{Offline Phase: Memory Profiling and Backdoor Design}\label{subsubsec:profiling}

The attacker must first characterize device-specific Rowhammer vulnerabilities and identify a set of target bits (in model weights) compatible with the profiled hardware faults.
We employ the Blacksmith tool~\cite{Blacksmith} to discover reproducible bit-flips on the selected DRAM. It employs non-uniform, frequency-domain Rowhammer patterns that vary the order and intensity of aggressor-row accesses, which bypass the Target Row Refresh (TRR) mitigation present in our commodity DDR4 modules. For each discovered bit-flip, we record the page frame number (PFN), the bit offset within the 4-KiB frame, the flip direction (whether the bit-flip is from 0 to 1 or from 1 to 0), and the specific Rowhammer pattern (including aggressor rows and the order of accessing them) that reliably reproduces the flip. Aggregating these records yields a \textit{fault map} indexed by PFN, which associates each vulnerable DRAM frame with its profiled Rowhammer pattern.

Given the fault map, BridgeHammer constructs a backdoored model tailored for the particular DRAM~\cite{robbin}. This is a tight software-hardware co-design process: all the bit-flips on chosen vulnerable DRAM frames constitute the backdoor. Compared to prior attacks that first select idealized target bit-flips for the model backdoor and then search for DRAM locations that can realize them (which inevitably may have unwanted collateral bit-flips), our hardware-tailored backdoor is more effective and strikes a better balance between ASR and CA.

The output of the offline phase is a set of target DNN pages-to-vulnerable DRAM frames mappings (reported in Section~\ref{sec:eval}) together with the complete profiled Rowhammer fault pattern associated with each vulnerable DRAM frame. These mappings and fault patterns drive the subsequent online model placement and fault-injection stages.

\subsection{Online Phase}\label{subsec:pipeline}

Figure~\ref{fig:main-fig} contrasts the standard DNN deployment pipeline on integrated platforms with the attacker-infected workflow. In the standard flow, the model serving environment first establishes a file-backed mapping of the model parameters (DNN pages) in CPU-managed physical memory, allocated by the OS. For GPU inference, the model is then compiled with supporting kernels (operations). The zero-copy weight anchoring binds the loaded model parameters into shared-memory tensors. Finally, input queries are staged and dispatched to the iGPU for inference. The adversary interrupts the workflow at two critical breakpoints, for precise memory massaging and hammering, respectively.

\subsubsection{\textbf{Breakpoint $T_1$} --  Targeted Memory Massaging}\label{subsec:massaging}

The first breakpoint occurs immediately after model compilation is completed and before zero-copy weight anchoring begins. The attacker monitors the victim's execution to detect this stage transition and temporarily suspends the victim process.

During the victim suspension time, the attacker aims to place the target DNN pages on vulnerable DRAM frames. As those pages have already been placed on physical frames during model loading, the attack must first release their current frames and then place the target DNN pages on specific vulnerable DRAM frames. We use an OS mechanism, the page cache, for releasing DNN pages. The page cache maintains file-backed pages in physical memory for faster access. The adversary induces system-wide memory pressure by using a large dummy buffer to continuously claim physical page frames. Both target and benign DNN pages may be evicted under the memory pressure. The attacker monitors the page-cache residency of the target DNN pages and terminates the eviction process once all target DNN pages have been evicted.

When the target DNN pages are released, we then perform normal memory massaging: curating the designated vulnerable DRAM frames according to the offline DNN-to-DRAM mapping in the per-CPU pageset (PCP), available for claiming in a last-in-first-out (LIFO) fashion. PCP keeps recently freed physical frames used by the Linux page allocator. The attacker touches the target DNN pages, as the associated target file offsets are known with file-backing, causing the kernel to draw memory frames from the PCP and allocate them to the DNN pages. When all target DNN pages are placed on corresponding vulnerable DRAM frames, the attacker resumes the victim with another signal \texttt{SIGCONT}. In the subsequent zero-copy binding process, victim accesses to the target DNN pages use the page-cache pages backed by the designated vulnerable frames. Benign DNN pages that were also evicted during the system-wide memory pressure are repopulated on demand through the normal kernel allocation path and do not require deterministic placement.

\subsubsection{\textbf{Breakpoint~$T_2$} -- Rowhammer-based Fault Injection}\label{subsec:injection} 

The second breakpoint occurs immediately after zero-copy weight anchoring has completed but before inference dispatch. The monitoring and interrupting mechanism is the same as for the first breakpoint. During the victim suspension time, the attacker performs Rowhammer. For each selected \textit{vulnerable DRAM frame}, the attacker activates the corresponding Rowhammer pattern recorded during offline profiling. This reproduces the profiled fault pattern for the frame. Once Rowhammer completes, the corruptions in \textit{target DNN pages} match exactly the physical fault patterns evaluated during the offline phase. When the victim resumes after the second breakpoint, inference is dispatched using the already-bound zero-copy tensors, causing the iGPU to access the corrupted model parameters in the shared memory.

\section{Attack Implementation on Intel Core i7-8700 }\label{sec:attack-inst}

We follow the methodology described in Section~\ref{sec:attack-meth} and implement a DNN backdoor attack. Our target platform is an 8th-generation Intel Core i7-8700 processor (Coffee Lake) integrated with Intel UHD Graphics 630 on the same die~\cite{intel-spec-update}. The CPU and iGPU share a single DRAM controller and a common pool of 8~GB DDR4 memory. We describe the zero-copy inference pipeline using the OpenVINO DNN inference framework, along with the Flush+Reload-based instruction monitoring and signaling that implement the breakpoints.

\subsection{Zero-Copy Model Deployment on CPU-iGPU Platforms}
\label{subsec:zero-copy-deploy}

On the CPU, the host uses \texttt{numpy.memmap} to load the model. OpenVINO then compiles the model for the UHD iGPU backend with \texttt{openvino.compile\_model}. The victim next performs zero-copy weight anchoring via unified memory management. The host application constructs OpenVINO Tensors over the file-backed memory with model parameters. By setting the \texttt{shared\_memory} attribute of these tensors to \texttt{true}~\cite{openvino-tensor}, OpenVINO lets the model access weights through existing host memory rather than copying them into separately allocated memory. The compiled executable graph is then explicitly bound to these shared-memory weight tensors. Finally, OpenVINO dispatches model inference to the iGPU by calling \texttt{infer()}.

\subsection{Page-Cache Eviction and PCP-based Placement}
\label{subsec:pcp-impl}

Traditionally, memory massaging happens during the CPU host-side model loading through PCP and OS-managed memory allocation~\cite{Deephammer}. However, PCP capacity limits the number of pages that can be staged for deterministic placement at a time~\cite{bolcskei2025rubicon}.

To implement the targeted memory massaging described in Section~\ref{subsec:massaging}, the attacker leverages file-backed model loading. The victim loads the DNN weights from a serialized model file, which is also accessible to the co-resident adversary. For each target DNN page for the backdoor, its page-aligned offset in the file is known. The attacker uses these file offsets to monitor their page-cache residency. Specifically, the attacker probes each target DNN page through its file offset using \texttt{preadv2()} with the configuration \texttt{RWF\_NOWAIT}. This allows the attacker to determine whether the file page under read remains resident in the page cache or not, without accessing the victim process's page tables. The attacker applies system-wide memory pressure until all selected target DNN pages have been evicted from the page cache.

For targeted placement through PCP, the attacker follows the
offline DNN-to-DRAM mapping and prepares the corresponding
vulnerable DRAM frame for each target DNN page. The attacker applies \texttt{madvise(MADV\_DONTNEED)} to its page backed by a vulnerable DRAM frame, returning the frame to the PCP. The attacker then immediately re-accesses a target DNN page with the page-aligned file offset, when the OS memory allocator draws the afore-prepared vulnerable DRAM frame from the PCP and places the target DNN page on it. The attacker repeats this release-and-reload procedure until all selected target DNN pages are placed on their corresponding vulnerable DRAM frames.

\begin{algorithm}[t]
\caption{Adversary process pseudo-code}
\small
\label{alg:alg1}
\begin{algorithmic}[1]
\Require Monitored instructions $I_1$ (return after \texttt{CompiledModel::create\_infer\_request}) and $I_2$ (return after \texttt{set\_input\_tensor}); 
timing threshold $\tau$; delay interval $\Delta$
\Ensure Malicious activities upon victim instruction execution 
\Loop
    \State \textsc{Flush} ($I_1$); \textsc{Flush} ($I_2$); \textsc{Delay} ($\Delta$)
    \State $t_1 \gets$ \textsc{TimeReload} ($I_1$) 
    \State $t_2 \gets$ \textsc{TimeReload} ($I_2$)
    \If{$t_1 < \tau$}
        \State \textsc{SIGSTOP}(victim)
        \State \textsc{EvictandMassage}()
        \State \textsc{SIGCONT}(victim)
    \EndIf

    \If{$t_2 < \tau$}
        \State \textsc{SIGSTOP}(victim)
        \State \textsc{Rowhammer}()
        \State \textsc{SIGCONT}(victim)
    \EndIf
\EndLoop
\end{algorithmic}
\end{algorithm}

\subsection{Flush+Reload-based Monitoring for Breakpoints}
A key requirement for the on-line attack phase is that the adversary must synchronize with the victim process to intercept it at the exact right time with malicious coordinated activities including memory massaging and Rowhammer. We use shared-library instruction monitoring to identify suitable locations for the two breakpoints.

Flush+Reload~\cite{zhang2016return} is a primitive for microarchitectural side-channel attacks, utilizing a special cache-management instruction, e.g., \texttt{CLFLUSH} on X86 architectures. Given the virtual address of an instruction, it is first flushed from any cache in the memory hierarchy. Later, when it is reloaded, we measure the execution latency. The delay is long due to a cache miss if the instruction is not executed since the last flush, but short (cache hit) if it has been accessed. The adversary chooses a critical instruction and repeatedly executes Flush-delay-Reload until the reload time falls below a preset threshold, indicating that the victim has just executed the instruction. Section~\ref{subsec:bp-sync} evaluates the selected observation window and the resulting synchronization margins.

We choose two victim instructions to monitor for interrupts,  i.e., anchors for the two breakpoints. The first one is the caller-side instruction immediately following \texttt{CompiledModel::create\_infer\_request()} in the OpenVINO \texttt{\_pyopenvino} shared object. It occurs after model compilation and inference-request creation but before the shared-memory weight-binding loop. The second one is the instruction immediately following \texttt{set\_input\_tensor()}, after zero-copy weight anchoring has completed but before inference dispatch. Algorithm~\ref{alg:alg1} shows the pseudocode for the adversary process, which monitors instructions and signals the victim process to interrupt.

\section{Evaluation}\label{sec:eval}
 
We evaluate backdoor effectiveness, runtime overhead, and synchronization reliability on the Intel platform with Ubuntu 22.04. We evaluate ResNet-20, VGG-16, and ResNet-50 on CIFAR-10 under both INT8 and FP32 precision. All relevant code and a detailed user manual for replicating the experiments in this work are available at \url{https://github.com/Yufei-Wang99/BridgeHammer}.

\begin{table}[t]
\centering
\caption{Backdoor results, model and target page counts, and bit-flip counts across the evaluated configurations.}
\label{tab:backdoor-results}
\footnotesize
\setlength{\tabcolsep}{3pt}
\begin{tabular*}{\columnwidth}{@{\extracolsep{\fill}}llrrrrrr@{}}
\toprule
Model & Quant. & \shortstack{Model\\pages} & \shortstack{Target\\pages} & \shortstack{Bit-\\flips} & \shortstack{Baseline\\Accuracy} & \shortstack{Clean\\Accuracy} & ASR \\
\midrule
ResNet-20 & INT8 & 66 & 18 & 70 & 90.5\% & 82.8\% & 89.7\% \\
VGG-16 & INT8 & 3657 & 130 & 319 & 92.0\% & 91.5\% & 90.1\% \\
ResNet-50 & INT8 & 6599 & 342 & 1513 & 90.9\% & 87.3\% & 84.4\% \\
\midrule
ResNet-20 & FP32 & 265 & 52 & 359 & 91.3\% & 85.0\% & 90.0\% \\
VGG-16 & FP32 & 14905 & 78 & 496 & 93.3\% & 85.9\% & 90.2\% \\
ResNet-50 & FP32 & 23678 & 250 & 1701 & 96.1\% & 85.1\% & 88.7\% \\
\bottomrule
\end{tabular*}
\vspace{-0.25cm}
\end{table}

Table~\ref{tab:backdoor-results} reports the attack results. Across the six configurations, the attack ASR ranges from 84.4\% to 90.2\%, while the resulting clean accuracy ranges from 82.8\% to 91.5\%. It also includes the backdoor attack complexity, including the target-page count against the model footprint and total number of bit-flips for each configuration. The evaluated models range from ResNet-20 INT8 with 66 pages to ResNet-50 FP32 with 23,678 pages, while the selected target DNN pages range from 18 to 342. Following the prior backdoor attack~\cite{robbin}, our bit-flip count represents the complete profiled fault patterns associated with the selected vulnerable DRAM frames. Across the evaluated configurations, these fault patterns contain 70--1,701 bit-flips. 

\begin{figure}[t]
\centering
\includegraphics[width=\linewidth]{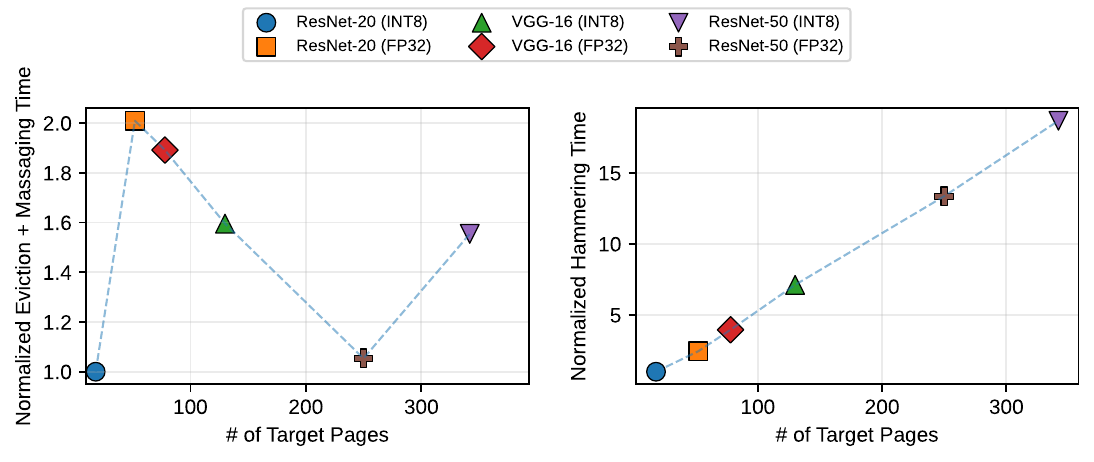}
\vspace{-0.5cm}
\caption{Normalized attack overhead across the evaluated configurations. (a) Eviction and massaging times at $T_1$, and (b) Rowhammering time at $T_2$. Both are normalized to ResNet-20 INT8.}
\label{fig:attack-overhead}
\vspace{-0.5cm}
\end{figure}

\subsection{Physical Attack Realization and Overhead}
\label{subsec:real-overhead}
The evaluated models span a wide range of memory footprints, while only the target DNN pages selected by the offline phase require deterministic physical memory placement. On our platform, we measure the PCP capacity at 1,135 4-KiB pages. The VGG-16 and ResNet-50 model footprints then exceed the PCP capacity, while BridgeHammer still performs the required targeted placement through our memory massaging technique with controlled evictions. 

Fig.~\ref{fig:attack-overhead} reports the normalized runtime cost of the two online attack stages. At breakpoint $T_1$, the delay overhead includes page-cache evictions followed by targeted PCP-based memory massaging. Because the memory pressure used for page-cache eviction is system-wide and may evict both target and non-target DNN pages in an interleaving way, the eviction time depends on how long it takes for all target DNN pages to be evicted, not necessarily linear to the target page count. At breakpoint $T_2$, the hammering time increases linearly with the number of target pages (DRAM frames to hammer).

\subsection{Synchronization and Reliability}
\label{subsec:bp-sync}

\begin{figure}[t]
\centering
\includegraphics[width=\columnwidth, trim=3 0 0 0, clip]{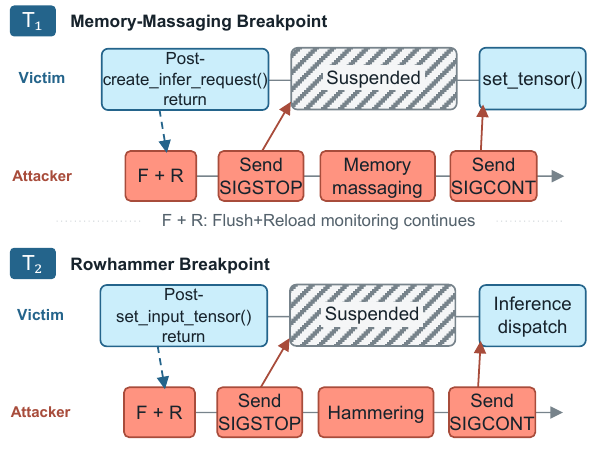}
\caption{Victim execution and attacker-side synchronization at $T_1$ and $T_2$.}
\label{fig:bp-overhead}
\end{figure}

Figure~\ref{fig:bp-overhead} illustrates the synchronization sequence between the adversary and the victim. We evaluate whether Flush+Reload monitoring can reliably detect the victim's execution progress and provide sufficient time to suspend the victim before \texttt{set\_tensor()} at $T_1$ and inference dispatch at $T_2$. We first characterize the Flush+Reload (F+R) monitor. On the evaluated platform, the cache hit latency is 44 cycles and cache miss is 214 cycles. We set the classification threshold at 120 cycles and use a 6.4K-cycle observation window between F+R. %

We next evaluate whether the attacker can suspend the victim in time. Across both breakpoints, the measured latency from detecting the monitored instruction to issuing \texttt{SIGSTOP} ranges from 7.55 to 25.51~$\mu$s. At $T_1$, all trials across the six evaluated model configurations suspend the victim before the first \texttt{set\_tensor()} call, with 84--296~$\mu$s of victim execution remaining. At $T_2$, the victim is suspended before inference dispatch, with 63.9--114.1~$\mu$s remaining before the subsequent \texttt{infer()} call. These results show that the selected monitoring points provide sufficient timing margins for reliable victim suspension at both breakpoints. The experiments run the victim and attacker alongside the default OS background services. On a different platform, the Flush+Reload threshold and synchronization timing should be re-characterized.

\section{Discussion and Conclusion}\label{sec:disc}

\noindent\textbf{Countermeasures.} Disabling zero-copy inference would sacrifice its performance benefits, so defenses should instead protect model-weight integrity within the shared-memory execution path. Potential approaches include pre-inference integrity verification, Rowhammer-aware memory allocation that excludes vulnerable physical pages, and monitoring for PCP-based memory-massaging activity.

To evaluate a practical software-level defense, we implement a runtime weight-integrity check that uses a reference SHA-256 digest generated offline and recomputes the digest over the weight buffers, after zero-copy weight anchoring and immediately before inference dispatch. A digest mismatch aborts the inference request before the iGPU consumes the corrupted weights. The pre-inference verification is performed once during model deployment, and its time complexity is linear in the total model-weight footprint. In our prototype, the median time overhead is 2.42~ms for ResNet-20 INT8 with 262~KiB of weights, 111~ms for VGG-16 INT8 with 14.3~MiB of weights, and 206~ms for ResNet-50 INT8 with 24.3~MiB of weights. The results demonstrate that persistent weight corruption introduced during the attack lifecycle under evaluation can be detected while preserving the existing zero-copy shared-memory execution path.

\noindent\textbf{Extension to Other Platforms.} We tested one Intel CPU--iGPU platform. Future work will examine NVIDIA Jetson (Arm CPU and GPU sharing LPDDR, TensorRT/CUDA~\cite{Jetson, jetson-xavier-nx-datasheet, cudaprogrammingguide}) and AMD Ryzen APUs (Zen CPUs, GPUs, XDNA NPUs, Ryzen AI with ONNX Runtime~\cite{amd_ryzen_ai_2024, amd_xdna_2023}).

\noindent\textbf{Conclusion.} BridgeHammer is the first cross CPU--iGPU Rowhammer backdoor against zero-copy DNN inference. On Intel hardware, it achieved an average ASR of 88.9\% and an average clean accuracy of 86.3\% across six configurations. These results motivate protecting model weights in shared-memory inference.

\clearpage
\bibliographystyle{IEEEtran}
\bibliography{ref}

\end{document}